\documentclass{article}
\usepackage{arxiv}
\usepackage{amsmath}
\usepackage{amssymb}
\usepackage{hyperref}
\usepackage{booktabs}  
\usepackage{bbm}
\usepackage{bm}
\usepackage{float}
\usepackage[numbers]{natbib}
\usepackage{graphicx}
\title{baskexact: An R package for analytical calculation of basket trial operating characteristics}
\author{
	Lukas Baumann \\
	Institute of Medical Biometry \\
	University of Heidelberg \\
	69120 Heidelberg, Germany \\
	\texttt{baumann@imbi.uni-heidelberg.de}
}

\begin{document}
\maketitle

\begin{abstract}
Basket trials are a new type of clinical trial in which a treatment is investigated in several subgroups. For the analysis of these trials, information is shared between the subgroups based on the observed data to increase the power. Many approaches for the analysis of basket trials have been suggested, but only a few have been implemented in open source software packages. The R package baskexact facilitates the evaluation of two basket trial designs which use empirical Bayes techniques for sharing information. With baskexact, operating characteristics for single-stage and two-stage designs can be calculated analytically and optimal tuning parameters can be selected. 
\end{abstract}

\section{Motivation and significance}

\subsection{Introduction}

Basket trials are a relatively new type of clinical trial in which a new treatment is tested in different subgroups. They are most commonly used in uncontrolled early-phase oncology trials, where a binary endpoint such as tumour response is investigated. The subgroups, also called baskets, are usually defined based on the tumour histology. All patients in the trial typically share a common genetic feature that is targeted by the treatment under investigation \citep{hirakawa2018}. Many designs for the analysis of basket trials have been proposed, often using Bayesian techniques such as Bayesian hierarchical modelling or Bayesian model averaging. By utilising these methods, information is shared between the subgroups based on the observed data, which is the defining characteristic of a basket trial from a statistical point of view \citep{pohl2021}. By sharing information, power is increased as compared to a separate analysis in each basket. This is essential, as sample sizes in the subgroups are often small due to their definition based on a certain genetic feature and tumour histology.

In their comprehensive review, \citet{pohl2021} categorised around 20 different basket trial designs and further designs were published in the meantime. Since many basket trial designs are methodology complex, reliable software implementation is vital for a design to be considered by researchers. However, only a fraction of the published designs are implemented in open source software \citep[see][for an overview]{meyer2021}. Searching the package list of the Comprehensive R Archive Network (CRAN) for the term "basket trial" only results in three hits (besides \texttt{baskexact}, the package presented in this article). \texttt{basket} \citep{basket1, basket2} implements the multisource exchangeability model design \citep{hobbs2018}. \texttt{bhmbasket} \citep{bhmbasket} implements two designs based on Bayesian hierarchical modelling, one by \citet{berry2013} and the EXNEX design by \citet{neuenschwander2016}. \texttt{bmabasket} \citep{bmabasket} implements a design based on Bayesian model averaging by \citet{psioda2021}. These methods are computationally relatively complex and thus operating characteristics are evaluated by simulation in these R packages. The methods implemented in \texttt{bhmbasket} additionally require Markov Chain Monte Carlo sampling to calculate posterior probabilities as the posterior distributions are not available in closed form when Bayesian hierarchical modelling is applied.

In this paper, the R package \texttt{baskexact} \citep{baskexact}, which implements the power prior basket trial design \citep{baumann2024} and a design by \citet{fujikawa2020} (henceforth referred to as Fujikawa's design) is presented. Both designs share information between baskets in a similar way using empirical Bayes methods. The power prior design is based on the power prior methodology which was initially proposed to borrow strength from historical data \citep{ibrahim2000}. Since both designs are not fully Bayesian, computation of posterior distributions is much cheaper than that of many other basket trial design. Posterior probabilities are available in closed form and can therefore easily be calculated even for a large number of baskets. Even analytical computation of operating characteristics is feasible in some settings. \texttt{baskexact} facilitates analytical computation of operating characteristics for single-stage and two-stage designs in basket trials with the power prior and Fujikawa's design, with equal sample sizes per subgroup and up to 5 baskets.

\subsection{Statistical background}

An uncontrolled basket trial with a binary endpoint and $K \geq 2$ baskets is considered. Let $n_k$ be the sample size in basket $k$. The number of responses in basket $k$ is denoted by $r_k$, the vector of responses in all baskets is denoted by $\bm{r}$. The objective of the trial is to identify the baskets in which the response probabilities, denoted by $p_k$, $0 \leq k \leq K$, is larger than a null response rate $p_0$, that is of no clinical interest. Thus, the null hypotheses $H_{0,k}: p_k \leq p_0$ are tested against the alternatives $H_{1,k}: p_k > p_0$.

In the power prior design, at first a beta prior with parameters $s_{1,k}$ and $s_{2,k}$ is specified for each baskets. Decisions are based on the following posterior distributions, which share the information between baskets:
\begin{equation*}
	\pi(p_k|\bm{r}, \bm{\omega_k}) = \text{Beta}(
	s_{1_k} + \sum_{i = 1}^K \omega_{k,i} r_i, 
	s_{2,k} + \sum_{i = 1}^K \omega_{k,i} (n_i - r_i)
	)
\end{equation*}
where $\omega_{k,i} \in [0, 1]$ are the sharing weights and $\bm{\omega_k} = (\omega_{k,1}, \dots, \omega_{k,K})$, with $\omega_{k,k} = 1$. $\omega_{k,i}$ is thus the proportion of information from basket $i$ that is used in the analysis of basket $k$. A null hypothesis is rejected if $P(p_k > p_0|\bm{r}, \bm{\omega_k}) \geq \lambda$, where $\lambda \in (0, 1)$ is prespecified.

The information sharing mechanism in the power prior design is very similar to that of Fujikawa's design. The difference between the sharing approaches in the two designs is solely that in Fujikawa's design prior parameters are also included in the weighted sums, and thus the prior information is also considered for the information sharing.

The sharing weights can be computed in various ways. In Fujikawa's design, they are derived from the pairwise Jensen-Shannon divergences (JSD) between the individual posterior distributions of the baskets in the following way:
\begin{equation*}
	\omega_{k,i} = \begin{cases}
		(1 - \text{JSD}(\pi(p_k|r_k), \pi(p_i|r_i))^\varepsilon & \text{if } (1 - \text{JSD}(\pi(p_k|r_k), \pi(p_i|r_i))^\varepsilon > \tau, \\
		0 & \text{otherwise},
	\end{cases}
\end{equation*}
where $\pi(p_k,r_k)$ is the individual posterior distribution of basket k (derived from a beta-binomial model) and $\varepsilon$ and $\tau$ are tuning parameters. Weights based on the JSD can of course also be used in the power prior design, i.e. without sharing the prior information as in Fujikawa's design.

Another way to calulcate the weights is derived from the so-called calibrated power prior (CPP) approach \citep{yuan2017}. CPP weights are based on the Kolmogorov-Smirnov test statistic, which for binary data is simply the absolute difference in response rates. CPP weights are calculated as follows:
\begin{equation*}
	\omega_{k,i} = \frac{1}{1 + \exp(a + b \log(d_{k,i} \max(n_k, n_i)^{1/4}))},
\end{equation*}
where $d_{k,i}$ denotes the absolute difference in response rates between basket $k$ and basket $i$.

In a two-stage design, an interim analysis is performed after a certain number of observations $n_{1,k}$ is available in each basket. \citet{fujikawa2020} suggest interim analyses based on the posterior predictive probability, which is the probability (based on prior information and interim data) that a result for which the null hypothesis can be rejected will be observed at the end of the study. If this probability is below or above a certain prespecified level in basket $k$, then that basket is stopped for futility or efficacy. Interim decisions may also be based on the same posterior distribution that is used for the final analysis.

For details about the two designs, the reader is referred to \citet{baumann2024} and \citet{fujikawa2020}.

\section{Software description}

\subsection{Software architecture}

\texttt{baskexact} is an R package, available on CRAN \citep{baumann2024} and GitHub (\url{https://github.com/lbau7/baskexact}). It is written using R's S4 and S3 methods. \texttt{baskexact} introduces two S4 classes, \texttt{OneStageDesign} and \texttt{TwoStageDesign}, corresponding to a single-stage and a two-stage basket trial design, respectively. This enables easy extension of the package. A vignette is provided which explains how \texttt{baskexact} can be extended.

To reduce computation times, some internal functions are implemented in C++, utilising the R packages Rcpp \citep{Rcpp} and RcppArmadillo \citep{RcppArmadillo}. The doFuture package \citep{doFuture, doFuture2} is used to enable parallelisation of some functions. With doFuture only a single function call is necessary to initialise parallelisation. The parallel backend can be chosen by the user.

\subsection{Software functionalities}

\texttt{baskexact} facilitates the evaluation of a basket trial with the power prior design and Fujikawa's design. It enables the analytical computation of the operating characteristics and selection of optimal tuning parameter values for single-stage and two-stage basket trials with equal sample sizes in all baskets. The most important functions are listed in Table \ref{tab:functions}. \texttt{baskexact} provides functions to compute basic operating characteristics such as the type 1 error rate (TOER), power and the expected sample size. In basket trials, also the expected number of correct decisions (ECD) is important as it in a sense combines TOERs and power of all baskets to a single number which is useful for comparing different designs and tuning parameter values.

\begin{table}[H]
	\centering
	\caption{Main functions of \texttt{baskexact}}
	\label{tab:functions}
	\begin{tabular}{@{}ll@{}}
		\toprule
		\textbf{Function} & \textbf{Description}                      \\ \midrule
		\texttt{toer}     & computation of family-wise and basket-wise TOERs \\
		\texttt{pow}      & computation of family-wise and basket-wise power \\
		\texttt{ecd}      & computation of expected number of correct decisions \\
		\texttt{ess}      & computation of expected sample size \\
		\texttt{estim}    & computation of mean posterior means and mean squared errors \\ 
		\texttt{adjust\_lambda} & find $\lambda$ that protects FWER at a certain level \\ 
		\texttt{opt\_design} & find optimal tuning paramter values based on ECD \\
		\texttt{plot\_weights} & plot a weight function \\
		\bottomrule
	\end{tabular}
\end{table}

To select the type of weights that are used for information sharing, a weight function has to be passed to the argument \texttt{weight\_fun} in each function. The key weight functions that can be selected are \texttt{weights\_cpp}, for the power prior design with CPP weights and \texttt{weights\_fujikawa} for Fujikawa's design, i.e. with weights based on the JSD. Both weight functions have additional tuning parameters which are passed as a list to the argument \texttt{weight\_params}. With the the function \texttt{opt\_design} the optimal tuning parameter values from a grid of values can be selected. Optimisation is based on the mean ECD across one or several scenarios, while protecting the family-wise TOER (FWER) under the global null hypothesis at a certain level.

In a two-stage design, additionally the type of interim analysis has to be specified using the argument \texttt{interim\_fun}. Available options are \texttt{interim\_postpred} for an interim analysis based on the posterior predictive probability and \texttt{interim\_posterior} for an interim analysis based on the posterior probability. To specify the probability stopping boundaries, both functions have two arguments: \texttt{prob\_futstop} and \texttt{prob\_effstop} for the futility and efficacy probability boundaries, respectively. These are passed as a list to \texttt{interim\_params}. Note that by setting \texttt{prob\_effstop} to 1, a design that only allows stopping for futility can be implemented.

\section{Illustrative examples}

The first step to use \texttt{baskexact} is always to create a design object corresponding either to a single-stage or a two-stage design (i.e. with one interim analysis) using the functions \texttt{setupOneStageBasket} and \texttt{setupTwoStageBasket}, respectively. For example:

\begin{verbatim}
> library(baskexact)
> design <- setupTwoStageBasket(k = 3, shape1 = 1, shape2 = 1, p0 = 0.2)
\end{verbatim}

\texttt{k} corresponds to the number of baskets, \texttt{shape1} and \texttt{shape2} to the two prior parameters of the beta prior distribution. Note that in \texttt{baskexact} equal priors for all baskets are assumed. \texttt{p0} refers to the response probability under the null hypothesis. The design object is the first argument of most function calls in \texttt{baskexact} to select the appropriate S4 method. For example, to compute the TOER:

\begin{verbatim}
> toer(
>   design = design,
>   n = 20,
>   n1 = 10,
>   lambda = 0.95,
>   interim_fun = interim_postpred,
>   interim_params = list(prob_futstop = 0.1, prob_effstop = 0.9),
>   weight_fun = weights_cpp,
>   weight_params = list(a = 1, b = 1),
>   results = "group"
> )

$rejection_probabilities
[1] 0.0569416 0.0569416 0.0569416

$fwer
[1] 0.1181975
\end{verbatim}

\texttt{toer} returns the basket-wise as well as the family-wise TOERs. \texttt{n} and \texttt{n1} refer to the maximum total sample size per basket and the sample size per basket for the interim analysis, respectively. \texttt{lambda} corresponds to the probability threshold $\lambda$ that is used in the final analysis to determine whether a null hypothesis is rejected. \texttt{interim\_fun} and \texttt{interim\_params} define the type of interim analysis. With \texttt{interim\_postpred}, interim decisions are based on the posterior predictive probability to reach a significant result at the end of the trial. \texttt{prob\_futstop = 0.1} and \texttt{prob\_effstop = 0.9} define that a basket is stopped for futility if this probability is below 0.1 and is stopped for efficacy if this probability is above 0.9. The other functions for the calculation of operating characteristics,  \texttt{pow}, \texttt{ecd}, \texttt{ess} and \texttt{estim}, work analogously.

In the planning stage of a basket trial, it may be desirable to select the probability threshold $\lambda$, such that the FWER is controlled at a certain level. This can be achieved with \texttt{adjust\_lambda}, which finds $\lambda$ such that the one-sided FWER under the global null hypothesis is smaller than a certain level \texttt{alpha}.

\begin{verbatim}
> adjust_lambda(
>   design = design,
>   alpha = 0.05,
>   n = 20,
>   n1 = 10,
>   interim_fun = interim_postpred,
>   interim_params = list(prob_futstop = 0.1, prob_effstop = 0.9),
>   weight_fun = weights_cpp,
>   weight_params = list(a = 1, b = 1),
>   prec_digits = 3
> )

$lambda
[1] 0.982

$toer
[1] 0.04807536
\end{verbatim}

Most arguments are the same as in \texttt{toer}. The only new arguments are \texttt{alpha} and \texttt{prec\_digits}. The latter specifies the number of decimal places of \texttt{lambda}. As the outcomes are binary, in general there is no $\lambda$ such that the FWER is exactly $\alpha$, as is seen in the output. Sometimes increasing \texttt{prec\_digits} results in a FWER closer the nominal level. Note that changing any of the design parameters (tuning parameters of the weight function or of the interim function) affects the FWER, i.e. the value for \texttt{lambda} is only valid for the given set of parameter values.

With \texttt{opt\_design}, tuning parameter values are sorted by their performance in terms of the mean ECD across a set of scenarios. A default set of scenarios can be created using \texttt{get\_scenarios}:

\begin{verbatim}
> get_scenarios(design = design, p1 = 0.5)
	
     0 Active 1 Active 2 Active 3 Active
[1,]      0.2      0.2      0.2      0.5
[2,]      0.2      0.2      0.5      0.5
[3,]      0.2      0.5      0.5      0.5
\end{verbatim}

This creates a matrix of response probability scenarios with an increasing number of baskets that are truly active. The null response probability is taken from the \texttt{design} object and \texttt{p1} specifies the response probability of the active baskets. Of course this set of scenarios can be extended or modified, e.g. to include scenarios in which active baskets have different response probabilities.

In \texttt{opt\_design}, for each combination of tuning parameter values that is passed to \texttt{weight\_params}, at first $\lambda$ is selected to control the FWER at a certain level and then, using the selected posterior threshold, the ECD is calculated for each scenario. Tuning parameters are then sorted by their mean ECD, where the mean over all scenarios is calculated. \texttt{opt\_design} can be run in parallel. For example, for a single-stage design:

\begin{verbatim}
> design <- setupOneStageBasket(k = 3, p0 = 0.2)
> plan(multisession, workers = 4)

> opt_design(
>   design = design,
>   n = 20,
>   alpha = 0.05,
>   weight_fun = weights_cpp,
>   weight_params = list(a = 1:3, b = 1:3),
>   scenarios = get_scenarios(design = design, p1 = 0.5),
>   prec_digits = 3
> )
	
a b Lambda 0 Active 1 Active 2 Active 3 Active Mean_ECD
1 2 1  0.981 2.932813 2.639612 2.636642 2.923344 2.783103
2 3 2  0.984 2.926667 2.655575 2.683766 2.859488 2.781374
3 3 3  0.983 2.928806 2.606198 2.661209 2.923073 2.779822
4 3 1  0.984 2.938167 2.703022 2.668577 2.803763 2.778382
5 2 2  0.978 2.919353 2.544335 2.590948 2.958013 2.753162
6 2 3  0.974 2.914952 2.438605 2.542111 2.976533 2.718050
7 1 1  0.973 2.917011 2.463110 2.468328 2.980259 2.707177
8 1 2  0.974 2.917205 2.365146 2.371869 2.989490 2.660927
9 1 3  0.971 2.888808 2.253843 2.360286 2.992850 2.623947
\end{verbatim}

Since the doFuture package is used for parallelisation, calling \texttt{plan} is enough to initialise evaluation in parallel. The code in the example makes all code run on 4 cores on the local machine. 

For each combination of tuning parameter values, the output contains the value for $\lambda$ which controls the FWER at the specified level, the ECD for each scenario and the mean ECD. The first row corresponds to the results of the tuning parameter values which achieve the highest mean ECD, the tuning parameter values are sorted in decreasing order by mean ECD.

\begin{figure}
	\centering
	\includegraphics[width=0.8\textwidth]{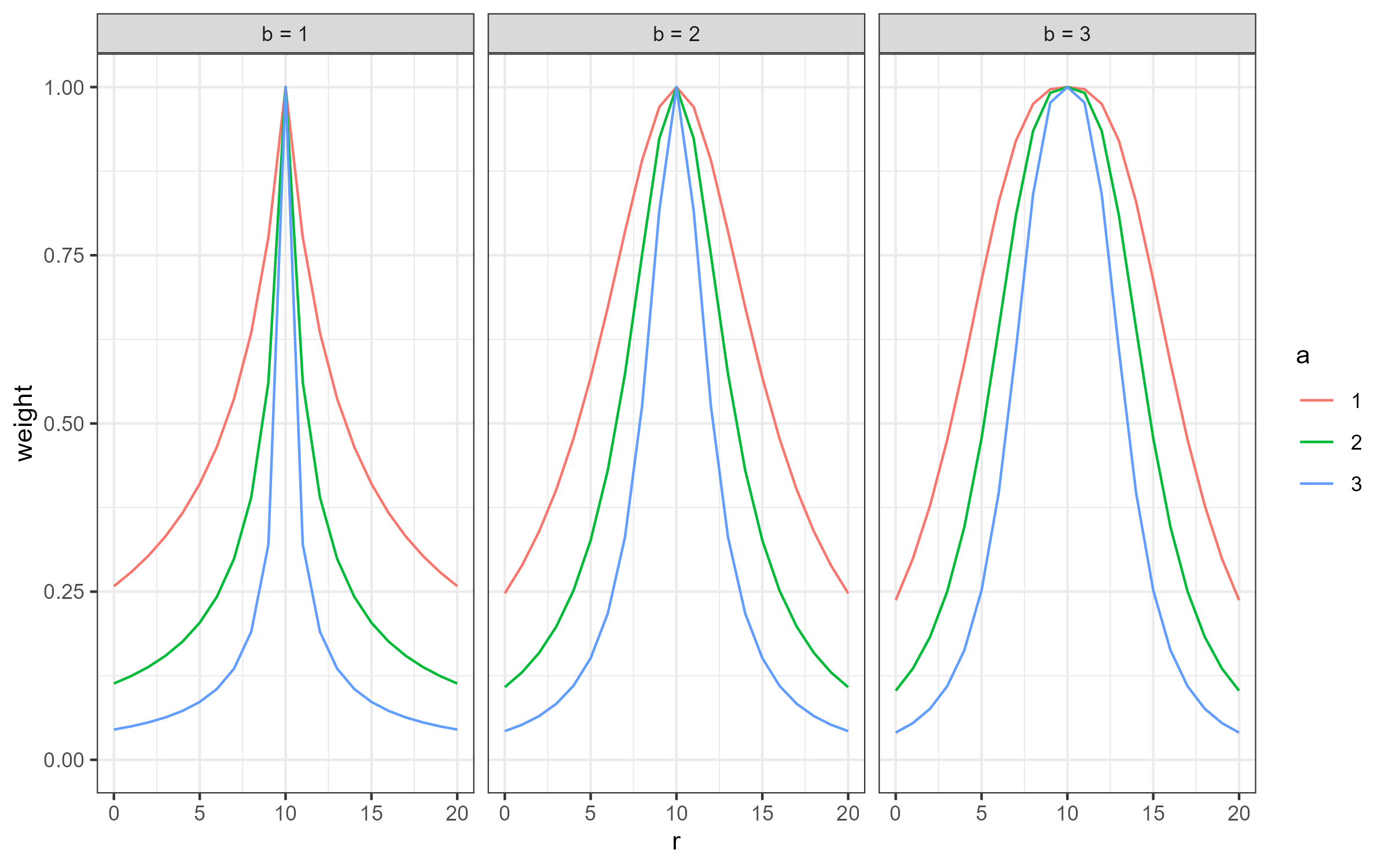}
	\caption{Weight function created with \texttt{plot\_weights}}
	\label{fig:weightplot}
\end{figure}

The amount of information that is shared between two baskets can be visualised with \texttt{plot\_weights}:

\begin{verbatim}
	> plot_weights(
	>   design = design, 
	>   n = 20, 
	>   r1 = 10, 
	>   weight_fun = weights_cpp, 
	>   weight_params = list(a = 1:3, b = 1:3)
	> )
\end{verbatim}

Figure \ref{fig:weightplot} shows the created plot. The argument \texttt{r1} refers to the number of responses observed in one of the two baskets. The responses in the other basket are varied on the x-axis. The weights resulting from different choices of the tuning parameters $a$ and $b$ are shown on the y-axis. As \texttt{r1} is set to 10, the weight is maximal when 10 responses are also observed in the other basket. The weights decline as the difference in responses increases, depending on the choice of tuning parameters values.

\section{Impact}

While many designs for the analysis of basket trials have been proposed in the literature, open source software implementation is the exception. However, without reliable, user-friendly and well documented software, it is unlikely that a design will be widely adopted. Software implementation is also necessary to enable comparison studies between different designs. Many basket trial designs were only evaluated in the manuscripts in which they were proposed. Thus, there is a large gap in the literature \citep{pohl2021}. Neutral comparison studies are necessary \citep{boulesteix2013} to determine which designs work best under which scenarios. Many questions regarding the application of basket trial designs are still open, e.g. the optimal type and timing of interim analyses. \texttt{baskexact} can thus be used
to investigate the performance of the power prior and Fujikawa's design. As the package is easily extendable, new weight functions for the power prior design can be implemented and evaluated which may further increase its impact.

\section{Conclusions}

In this manuscript, the R package \texttt{baskexact} was presented which implements two closely related basket trial designs, the power prior design and Fujikawa's design, which both share information using an empirical Bayes method and are therefore computationally much cheaper than other Bayesian basket trial designs. \texttt{baskexact} enables analytical computation of operating characteristics for up to 5 baskets and single-stage and two-stage designs. The main limitation of \texttt{baskexact} is that currently only equal sample sizes per basket are supported. Nevertheless, \texttt{baskexact} is a valuable tool for the evaluation of the two implemented designs and can be used by researchers to further compare the performance of different basket trial designs.

\section*{Acknowledgements}

Thanks to Marietta Kirchner for valuable comments on a first draft of this manuscript.

\bibliographystyle{unsrtnat}
\bibliography{references}

\end{document}